%% file: ms.tex
\pdfoutput=1

\documentclass[12pt,a4paper,notitlepage,oneside]{book}

\include{header-alleviushoehle}

\usepackage{abstract_def}

\usepackage{titling}
\usepackage{authblk}
\title{Prospective Detection of Outbreaks}
\author[1]{Benjamin Allévius}
\author[1]{Michael Höhle}
\affil[1]{Department of Mathematics, Stockholm University, Sweden}
\date{\vspace{-5ex}}

\begin{document}
\maketitle

\framebox{
  \begin{minipage}{0.9\linewidth}
    This manuscript is a preprint of a Chapter to appear in the
    \textit{Handbook of Infectious Disease Data Analysis}, Held, L.,
    Hens, N., O'Neill, P.D. and Wallinga, J. (Eds.). Chapman \& Hall/CRC,
    2018. Please use the book for possible citations.
  \end{minipage}
}\\[2ex]

\begin{chapabstract} 
  \input{abstract}  
\end{chapabstract}
\thispagestyle{empty}
\thispagestyle{empty}
\clearpage
\setcounter{page}{1}
\input{chapter-multivariate}
\bibliographystyle{apalike}
\bibliography{ms}
\end{document}

%% file: header-alleviushoehle.tex
\usepackage{alltt}
\usepackage[T1]{fontenc}
\usepackage[utf8]{inputenc}
\usepackage[english]{babel}

\usepackage{amsmath, amsthm, amssymb, mathtools}
\usepackage{dsfont}
\usepackage[cal=cm]{mathalfa}
\numberwithin{equation}{section}
\usepackage{chngcntr}
\counterwithout{equation}{section}

\usepackage{setspace} 
\usepackage{graphicx}
\usepackage{adjustbox}
\usepackage[usenames,dvipsnames]{xcolor}

\usepackage{float} 

\usepackage{booktabs}

\usepackage{todonotes} 
\usepackage{url}
\usepackage{pdflscape} 
\usepackage{framed} 
\usepackage{enumitem} 

\usepackage[iso,swedish]{isodate}

\usepackage[round,colon,authoryear]{natbib}

\renewcommand{\P}{\mathbb{P}} 
\newcommand{\E}{\mathbb{E}} 
\newcommand{\V}{\mathbb{V}}
\newcommand{\bs}{\boldsymbol}
\newcommand{\IX}{\bs{1}}

%% file: abstract.tex
\small{
\noindent 
This chapter surveys univariate and multivariate methods for
infectious disease outbreak detection. 
The setting considered is a prospective one: data arrives sequentially as part 
of the surveillance systems maintained by public health authorities, and the 
task is to determine whether to `sound the alarm’ or not, given the recent 
history of data. 
The chapter begins by describing two popular detection methods for univariate 
time series data: the EARS algorithm of the CDC, and the Farrington algorithm 
more popular at European public health institutions. This is followed by a 
discussion of methods that extend some of the univariate methods to a 
multivariate setting. This may enable the detection of outbreaks whose signal is 
only weakly present in any single data stream considered on its own. 
The chapter ends with a longer discussion of methods for outbreak detection in 
spatio-temporal data. These methods are not only tasked with determining 
\emph{if} and \emph{when} an outbreak started to emerge, but also \emph{where}. 
In particular, the scan statistics methodology for outbreak cluster detection in 
discrete-time area-referenced data is discussed, as well as similar methods for 
continuous-time, continuous-space data. 
As a running example to illustrate the methods covered in the chapter, a 
dataset on invasive meningococcal disease in Germany in the years 2002--2008
is used. This data and the methods covered are available through the R packages 
\texttt{surveillance} and \texttt{scanstatistics}.
}

%% file: chapter-multivariate.tex
\setcounter{chapter}{-1}

\chapter{Prospective Detection of Outbreaks}
\noindent
An essential aspect of infectious disease epidemiology is the timely
detection of emerging infectious disease threats. To facilitate such
detection, public health authorities maintain surveillance systems for
the structured collection of data in humans, animals and plants. In
this chapter we focus on the \textit{prospective}, i.e.\
as-data-arrive, detection of outbreaks in such data streams obtained
as part of the routine surveillance for known diseases, symptoms or
other well-defined events of interest. The organization of this
chapter is as follows: In Section~\ref{sec:univariate} we briefly
present methods for outbreak detection in purely temporal data
streams, which is followed by a more extensive presentation of methods for
spatio-temporal detection in Section~\ref{sec:multivariate}.

\section{Motivation and Data Example} \label{sec:motivation}

Surveillance data is nowadays collected in vast amounts to support the
analysis and control of infectious diseases. As a consequence, the
data volume, velocity and variety exceeds the resources to look at
each report individually and thus statistical summaries, insightful
visualizations and automation are needed~\citep{hoehle2017b}. In
response to this, semi-automatic systems for the screening and further
investigation of outbreaks have been developed. This screening
typically consists of identifying emerging spikes in the monitored
data streams and flagging particular cases for further inspection. For
foodborne diseases, for example, this inspection could consist of
attempts to identify and remove the food source consumed by all cases.

Our aim is thus to develop data mining tools to support the sequential
decision making problem (to react or not) for pathogens with quite
heterogeneous features, differing in e.g., data collection, prevalence
and public health importance. As an example, even a single Ebola case
constitutes a serious public health threat, whereas only larger
clusters of \textit{Campylobacter jejuni} infections will trigger
public health actions. It is also worth to point out that large
accumulations of cases in a short time period are almost surely
noticed at the local level. Automatic procedures nonetheless provide a
safety net ensuring that nothing important is missed, but the added
value lies predominantly in the detection of dispersed cross-regional
outbreaks. Another added value is that the quantitative nature of the
algorithms allows for a more objective approach, which can be a
helpful addition to epidemiological intuition. However, alarms are
useless if they are too frequent to be investigated properly. On the
other hand, missing an important outbreak is also fatal.

\subsection{Invasive Meningococcal Surveillance in Germany}

\begin{figure}[H]
  \centering
  \includegraphics[scale=0.55]{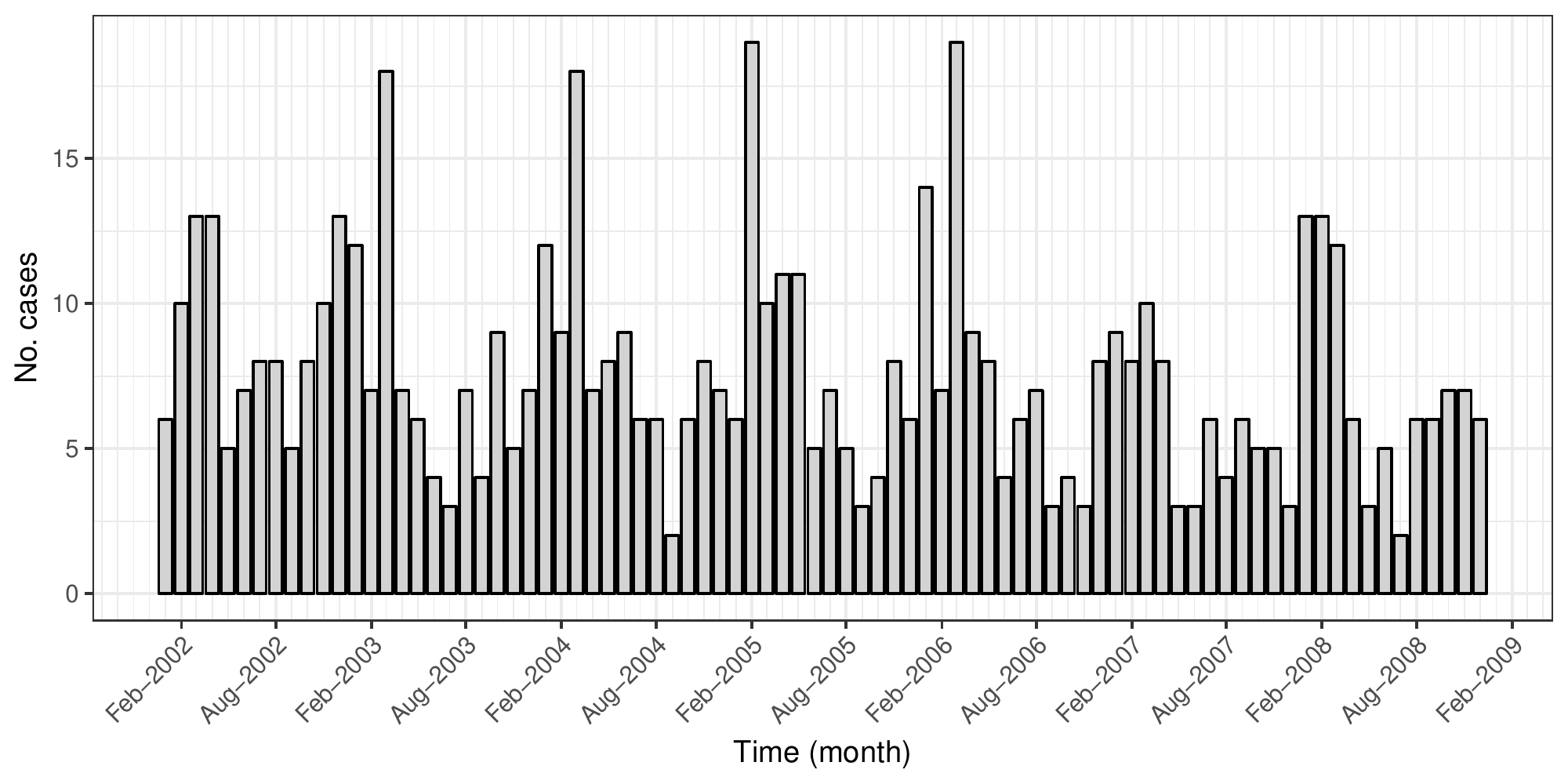}
  \caption{Time series of monthly number of IMD cases of the two
    finetypes.} \label{fig:meningo-monthly-ts}
\end{figure}

As an illustration we use 2002-2008 data from the routine monitoring
of invasive meningococcal disease (IMD) by the National Reference
Centre for Meningococci (NRZM) in
Germany\footnote{http://www.meningococcus.de} as motivating
example. Figure~\ref{fig:meningo-monthly-ts} shows the number of new
cases of two particular finetypes described in more detail in
\citet{Meyer2012} and available as dataset \texttt{imdepi} in the R
package \texttt{surveillance}~\citep{salmon_etal2016a,meyer_etal2017}.

From the figure and its context we observe a number features which
make the statistical modelling and monitoring of surveillance time
series a challenge.  From the time series in the figure we see that we
deal with a low integer number of reported cases, which can contain
secular trends, seasonality, as well as past outbreaks. In the IMD
example a microbiological investigation provides a pretty clear
definition, however, having a clear case definition can otherwise be a
challenge. Furthermore, it is common to only report positive test
outcomes with no information on how many tests were actually
performed. Spikes in the incidence of new cases might very well be due
to an increased testing behaviour.  This brings us to the important
issues of under-ascertainment and under-reporting often reflected as
stages in a surveillance pyramid of which only the top stage are the
reported cases. Another issue, which we shall ignore in this chapter,
is the problem of reporting delays; depending on the aims it can be
necessary to adjust the detection for this phenomena, c.f.\
the chapter written by Angela Noufaily. Instead, we move on to a
statistical description and solution of the detection problem.

\section{Univariate Surveillance Methods} \label{sec:univariate}

Assume data have been pre-processed such that a univariate time series
of counts is available. Denote by $y_1, \ldots, y_t$ the time series
of cases with the time index representing, e.g., daily, weekly or
monthly time periods and $y_t$ being the observation under present
consideration. Different statistical approaches can now be used to
detect an outbreak at the last instance $t$ of the series. Such
approaches have a long tradition and comprehensive review articles
exist~\citep{farrington_andrews2003, Unkel2012,
  Sonesson2003,lestrat2005,Woodall2006,hoehle_mazick2011}. An
implementational description of many of these algorithms can be found
in the R package \texttt{surveillance}~\citep{salmon_etal2016a}. We
therefore keep this account short and just present two of the most
commonly used algorithms, in particular because their approach fits
well into that of the multivariate approaches covered in
Section~\ref{sec:multivariate}. Both algorithms presented compute a
prediction for $y_t$ based on a set of historic values, under the
assumption that there is no outbreak at $y_t$, and then assess how
extreme the actually observed value is under this assumption.

\subsection{EARS Algorithm}

The Early Aberration Detection System (EARS) method of the CDC as
described in~\citet{SIM:SIM3197} is a simple algorithm convenient in
situations when little historic information is available and, hence,
trends and seasonality are not of concern. In its simplest form, the
baseline is formed by the last $k=7$ timepoints before the assessed
timepoint $t$, which is particularly meaningful when monitoring a time
series of daily data. The simplest version of the method is based on
the statistic $C_{t} = (y_{t}-\bar{y}_{t})/s_{t}$, where
\begin{align*}
\bar{y}_{t} = \frac{1}{7} \sum_{s=t-7}^{t-1} y_s \quad\text{and}\quad
s_{t}^2= \frac{1}{k-1} \sum_{s=t-k}^{t-1} \left(y_s -
  \bar{y}_{t}\right)^2
\end{align*}
are the usual unbiased estimators for the mean and variance of the
historic values. For the null hypothesis of no outbreak, it is assumed
that $C_{t} \sim {N}(0,1)$. The threshold for an extreme observation,
under the assumption of no outbreak, is now defined as $U_{t}=
\bar{y}_{t} + 3\cdot s_{t}$. Consequently, an alarm is raised if the
current observation $y_{t}$ exceeds this upper limit.

The simplicity of the method makes it attractive, and variants of this
Gaussian approach can be found in different contexts, e.g., as an
approximation when the underlying distribution is
binomial~\citep{andersson_etal2014}. However, from a statistical point
of view the method has a number of shortcomings. In particular the
distributional $N(0,1)$ assumption is likely to be inaccurate in case
of counts below 5-10, because the distribution is discrete,
non-negative and hence right-skewed. As easy as the three times
standard deviation is to remember, the appropriate comparison of $y_t$
should be with the upper limit of a prediction interval. Assuming that
the $y$'s are identical and independent variables from a Gaussian
distribution with both mean and standard deviation unknown, such a
one-sided $(1-\alpha)\cdot 100\%$ interval has upper limit
\begin{align} \label{eq:predict-ul-gauss}
  \bar{y}_{t} + t_{1-\alpha}(k-1) \cdot s_{t} \cdot
  \sqrt{1+\frac{1}{k}},
\end{align}
where $t_{1-\alpha}(k-1)$ denotes the $1-\alpha$ quantile of the
t-distribution with $k-1$ degrees of freedom. Note that for $k$ larger
than 20-30, this quantile is very close to that of the standard normal
distribution, which is why some accounts,
e.g.~\citet{farrington_andrews2003}, use $z_{1-\alpha}$
instead. However, when $k=7$ the $\alpha$ corresponding to a
multiplication factor of 3 is not $\alpha = 1-\Phi(3)= 0.0013$, as one
might think, but $\alpha=1-F(3/\sqrt{1+1/k})=0.0155$, where $F$
denotes the cumulative distribution function of the t-distribution
with $k-1$ degrees of freedom. In other words, using a factor of 3
means that the probability of false discoveries is notably higher than
one might naively expect. From an epidemiologist's point of view, this
might seem as yet another instance of statistical nitpicking, however,
in the next section we provide an example illustrating how misaligned
the $3\cdot s_t$ rule can be in the case of few and low count data.

The more historic values are available, the larger $k$ one can choose
in practice.  However, seasonality and secular trends then become an
issue. A simple approach to handle seasonality suggested
by~\citet{stroup_etal1989} is to just pick time points similar to the
currently monitored time point. For example: When working with weekly data
and looking at week 45 in year 2017 then one could take the values of,
say, weeks 43 to 47 of the previous years as historic values. Another
approach is to handle seasonality and trends explicitly in a linear
regression model framework, e.g.,
\begin{align*}
\mu_t = \E(y_t) = \beta_0 + \beta_1\cdot t + \sum_{l=1}^L \left\{ \beta_{2l}
\sin\left(\frac{2\pi l t}{P}\right) + \beta_{2l+1} \cos\left(\frac{2\pi l t}{P}\right)\right\},
\end{align*}
where $P$ is the period, e.g., 52 for weekly data. This has the
advantage that \textit{all} historic values are used to infer what is
expected. As an alternative to the above superposition of harmonics
one could instead use (penalized) cyclic splines~\citep{wood2006} or
factor levels for the individual months or days.

\subsection{Farrington Algorithm} \label{sec:farrington}

An extension of the above approaches is the so called Farrington
algorithm~\citep{farrington_etal1996,noufaily_etal2013}, which
explicitly uses an underlying count data distribution and handles
possible trends through the use of an (over-dispersed) Poisson
regression framework. The algorithm is particularly popular at
European public health institutions as its easy to operate and handles
a large spectrum of time series with different characteristics without
the need of particular tuning.

Assuming, as before, that one wants to predict the number of cases
$y_{t}$ at time $t$ under the assumption of no outbreak by using a set
of historic values from a window of size $2w+1$ up to $b$ periods back
in time. With weekly data and assuming 52 weeks in every year the set
of historic values would be $\cup_{i=1}^b \cup_{j=-w}^w\>\> y_{t-i
  \cdot 52 +j}$. We then fit an over-dispersed Poisson generalized
linear model (GLM) with $\V(y_s)=\phi\cdot\mu_s$  and log-linear predictor
\begin{align*}
 \E(y_s)=\mu_s, \quad\text{where}\quad \log(\mu_s) = \beta_0 + \beta_1 \cdot s
\end{align*}
to the historic values. In the above, $s$ denotes the $b(2w+1)$ time
points $t-i \cdot 52 + j$ of the historic values and $\phi>0$ is the
over-dispersion parameter. If the dispersion parameter in the
quasi-Poisson is estimated to be smaller than one a Poisson model
(i.e.\ $\phi=1$) is used instead. Based on the estimated GLM model we
compute the upper of limit of a $(1-\alpha)\cdot 100\%$ one-sided
prediction interval for $y_{t}$ by
\begin{align}
  \label{eq:farrington-pi}
 U_{t} &=\hat{\mu}_t + z_{1-\alpha}\cdot \sqrt{\V(y_{t} -
  \hat{\mu}_t)},
\end{align}
where $\V(y_{t} - \hat{\mu}_t) = \V(y_{t}) + \V(\hat{\mu}_t) -
2\operatorname{Cov}(y_{t},\hat{\mu_t}) = \phi\mu_t + \V(\hat{\mu}_t)$,
because the current observation is not used as part of the estimation.
We know that asymptotically $\hat{\bs{\beta}} \stackrel{a}{\sim}
N(\bs{\beta},I^{-1}(\bs{\beta}))$, where $I^{-1}$ denotes the inverse
of the observed Fisher information. Therefore,
$\hat{\eta}_t=\log(\hat{\mu}_t)$ is normally distributed with variance
$\V(\hat{\eta}_t)=\V(\hat{\beta}_0)+t^2 \V(\hat{\beta_1}) +
2t\operatorname{Cov}(\hat{\beta}_0,\hat{\beta}_1)$. Through the use of
the delta method we find that $\V(\hat{\mu_t}) \approx
\exp(\hat{\eta}_t) \V(\hat{\eta}_t)$.

\begin{figure}[H]
  \centering
  \includegraphics[scale=0.75]{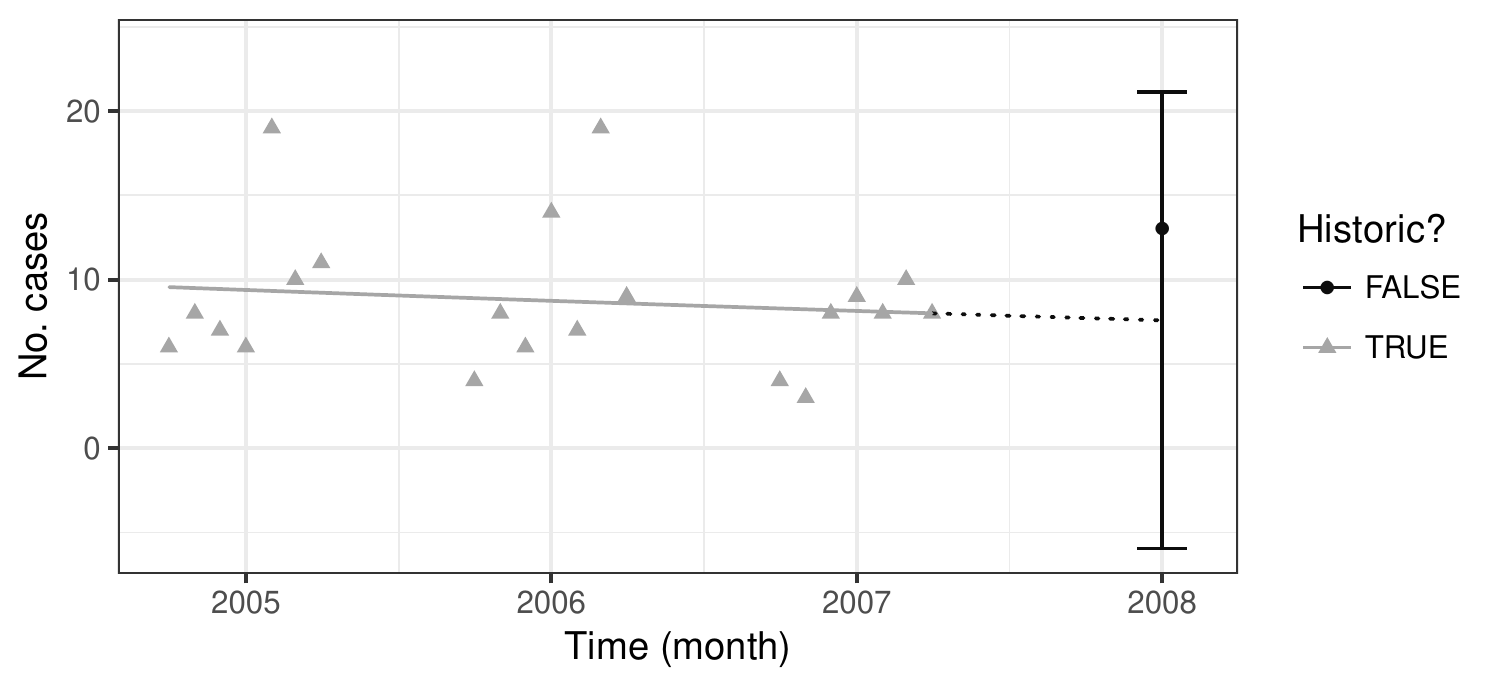}
  \caption{The historic values, fitted mean and prediction interval of
    the simplified Farrington procedure for the observation of January 2008.} \label{fig:farrington-predict71}
\end{figure}

Figure~\ref{fig:farrington-predict71} displays the fitted GLM and the
limits of a corresponding two-sided $(1-2\alpha)\cdot 100\%$
prediction interval for the first observation in 2008 when $b=3$,
$w=3$ and $\alpha=0.00135$. Because the upper limit 21.1 of the
prediction interval is larger than the observed value of 13 no alarm
is sounded.

Note that the above is a somewhat simplified description of the
Farrington procedure. For example, a trend is only included, if the
following three conditions are all fulfilled: it is significant at the
0.05 level, there are at least 3 years of historic values and
including it does not lead to an over-extrapolation (i.e.\
$\hat{\mu}_t$ is smaller than the maximum of the historic
values). Additional refinements of the algorithm include the
computation of the prediction interval in \eqref{eq:farrington-pi} on
a $\sqrt{y}$ or $y^{2/3}$ power transformation scale and then
back-transform the result. This would for example fix the obvious
problem in \eqref{fig:farrington-predict71} that the prediction
includes negative values: using the 2/3-power transformation the
resulting upper limit would be 24.5 instead. If the historic values
contain outliers in form of previous outbreaks the Farrington
algorithm suggests to instead base the prediction on a re-weighted
second fit with weights based on Anscombe
residuals. \citet{noufaily_etal2013} suggest a number of additional
improvements to the algorithm. Instead of using the window based
approach to select the historic values, all past data are utilized in
the regression model by adding a cyclic 11-knot zero-order spline
consisting of a 7-week historic period and nine 5-week
periods. Furthermore, better results seem to be obtained when
computing the prediction interval directly on a negative binomial
assumption, i.e.\ by assuming $y_t \sim
\operatorname{NegBin}\left(\mu_t,\mu_t/(\phi-1)\right)$ where the
first argument denotes the mean and the second argument the dispersion
parameter of the distribution. By plug-in of the estimates
$\hat{\mu}_t$ and $\hat{\phi}$ one obtains that the $1-\alpha$
quantile of the negative binomial distribution in the example of
Fig.~\ref{fig:farrington-predict71} is 24.

For comparison: had we taken the 7 observations corresponding to
Farrington's $b=1$ and $w=3$ as historic values for the EARS approach
we get $\overline{y}_t=7.1$, $s_t=2.6$, and an upper limit of
15.0. When using \eqref{eq:predict-ul-gauss} with $\alpha=1-\Phi(3)$
for the same historic values the upper limit is 20.8. The
corresponding upper limit of the Farrington procedure are 15.7
(untransformed), 17.2 (2/3-power transformation) and 16 (Poisson
quantile) for the Quasi-Poisson approach. If we assume that a
$\operatorname{Po}(7.1)$ is the correct null-distribution (the output
of the Farrington algorithm in the example), the 3 times standard
deviation rule thus produces too many false alarms---in this
particular setting about 0.681\% instead of the nominal
0.135\%. Continuing this further: If the 7 observations would instead
have been a quarter of their value (i.e.\ 2x1 and 5x2), the
corresponding false alarm probability of the $3\cdot s_t$ approach
would raise as high as 9.53\%, whereas the Farrington procedure with
quantile threshold by construction keeps the nominal level. This
illustrates the importance of using count response distributions when
counts are small.

\section{Multivariate Surveillance Methods} \label{sec:multivariate}
Surveillance of univariate data streams typically involves some type of spatial
or temporal aggregation. For example, many of the detection algorithms discussed
in the previous section assume the disease cases are counted on a daily or weekly
basis, or monitor just the total sum of all cases that occur in a municipality, a county,
or even a whole nation. If this aggregation is too coarse, it could result in the
loss of information useful for the detection of emerging threats to public health.
Similarly, the failure to monitor secondary data sources such as
over-the-counter medicine sales may equate to a forfeiture of the same. These
problems motivate the use of \emph{multivariate} methods, which offer means to
treat data at a finer spatial or temporal scale, or to include more sources of
information. We start this section by reviewing some of the available
extensions of univariate methods to multivariate settings, and then proceed to
cover methods for cluster detection in spatio-temporal data. Naturally, our
review cannot be all-encompassing, and we therefore direct the reader to the
books by \citet{Lawson2005}, \citet{Wagner2006}, and \citet{Rogerson2008}
for other takes on surveillance methods for spatio-temporal data.

\subsection{Extensions of Univariate Surveillance Methods}
An intuitive approach to the surveillance of multiple data streams is to apply
one of the univariate methods from Section \ref{sec:univariate}
to each time series monitored. For example, \citet{Hohle2009} applied the
Farrington method to the monthly incidence of rabies among
foxes in the 26 districts of the German state of Hesse, and did likewise to the
aggregated cases for the state as a whole and for its subdivision into three
administrative units. Data from the years 1990--1997 were available as a
baseline, and the period 1998--2005 was used for evaluation. The authors were
able to to pinpoint the districts in which an outbreak occurred in March of 2000.
\citet{Hohle2009} argue that using multiple univariate detectors in this
hierarchical way is often a pragmatic choice, because many of the analogous
multivariate changepoint detection methods used in the literature
\citep[see e.g.][]{Rogerson2004a} assume continuous distributions for the data;
an assumption hardly realistic for the low count time series often seen after
partitioning the total number of cases by region, age, serotype, and so on.

The parallel application of univariate methods does have its downsides, however.
Univariate methods such as the Farrington algorithm 
require either a false alarm probability (significance level) $\alpha$ or a
threshold $c$ to be set prior to an analysis. These are often set to achieve
some maximum number of false alarms per month or year
\citep[see e.g.][for other optimality criteria]{Frisen2003}.
If the same conventional $\alpha$ is used for $p \gg 1$ detection methods run in
parallel, in the absence of an outbreak, the probability of raising at
least one false alarm will be much greater than $\alpha$.
On the other hand, lowering  $\alpha$ will make outbreaks harder to detect.
Multivariate methods, considered next, do not suffer from the same issues.


\subsubsection{Scalar Reduction and Vector Accumulation}
In the multivariate setting, we suppose that the process under surveillance can
be represented as a $p$-variate vector $\bs{Y}_t = (Y_{t,1}, Y_{t,2}, \ldots,
Y_{t,p})'$, where $t = 1, 2, \ldots$ are the time points under consideration.
Each component $Y_{t,i}$ could represent the
disease incidence (as a count) of a given region at time $t$, for example. One
of the earliest \emph{control chart} methods of multivariate surveillance is the
use of \emph{Hotelling's $T^2$ statistic} \citep{Hotelling1947}.
Under the null hypothesis for this method, $\bs{Y}_t$ is assumed to follow a
multivariate normal distribution with mean vector $\bs{\mu}$ and covariance
matrix $\bs{\Sigma}$, both typically estimated from the data. Hotelling's method
then reduces the multivariate observation at each timepoint to a scalar
statistic, given by
\begin{align} \label{eq:hotellingT2}
  T^2_t = \left(\bs{Y}_t - \hat{\bs{\mu}} \right)'
           \hat{\bs{\Sigma}}^{-1}
           \left(\bs{Y}_t - \hat{\bs{\mu}} \right), \quad t = 1, 2, \ldots, n.
\end{align}
The Hotelling $T^2$ statistic is thus the squared Mahalanobis distance,
which measures the distance between the observed data and the null hypothesis
distribution while accounting for the different scales and correlations of the
monitored variables.
When properly scaled, 
the $T^2$ statistic has an
$F_{p,n-p}$-distribution under the null hypothesis of no outbreak; hence a
detection threshold is given by a suitable quantile from this distribution.
When dealing with disease count data, however, we know beforehand that the
$Y_{t,j}$s should \emph{increase} in case of an outbreak. This prior knowledge
is not reflected in the $T^2_t$ statistic, which penalizes deviations from
the mean in either direction. With this motivation, \citet{OBrien1984}
proposed several parametric and non-parametric tests that accomodate alternative
hypotheses for consistent departures from the null hypothesis.

A problem with Hotelling $T^2$ statistic, and likewise the methods proposed by
\citet{OBrien1984}, is that they will fail to accumulate evidence for an
outbreak over time.
As noted by \citet{Sonesson2005} (in regard to the $T^2$ statistic), this will
render the methods ineffective at detecting small to moderate changes in the
monitored process. A solution suggested by \citet{Crosier1988} is to first
calculate the Hotelling $T_t^2$ statistic for each timepoint $t$ in the
surveillance period, take its square root $T_t$, and then apply a CUSUM
\citep{Page1954} scheme
$S_t = \max\{0, S_{t-1} + T_t - k\}$, where $S_0 \geq 0$ and $k > 0$ are chosen
together with a threshold $c$ to achieve a certain false positive rate, for
example. Similarly, \citet{Rogerson1997} devised a CUSUMized version of
\citeauthor{Tango1995}'s \citeyearpar{Tango1995} retrospective spatial
statistic, which assumes a Poisson distribution for the data, and calculates a
quadratic form statistic using a distance matrix.

The parallel surveillance and scalar reduction methods can be combined
by first calculating the ordinary alarm statistics for each data stream, and
then apply a scalar reduction method to the vector of such statistics. A few
such approaches are reviewed in \citet{Sonesson2005}. We also point the reader
to the important review papers by \citet{Frisen1992}, \citet{Sonesson2003},
\citet{Frisen2010}, and \citet{Unkel2012}. Finally, for a more general
introduction to statistical quality control and the CUSUM method in particular,
we recommend the book written by \citet{Montgomery2008}.

\subsubsection{Illustration of Hotelling's $T^2$ statistic}
\begin{figure}[H]
  \centering
  \includegraphics[scale=0.75]{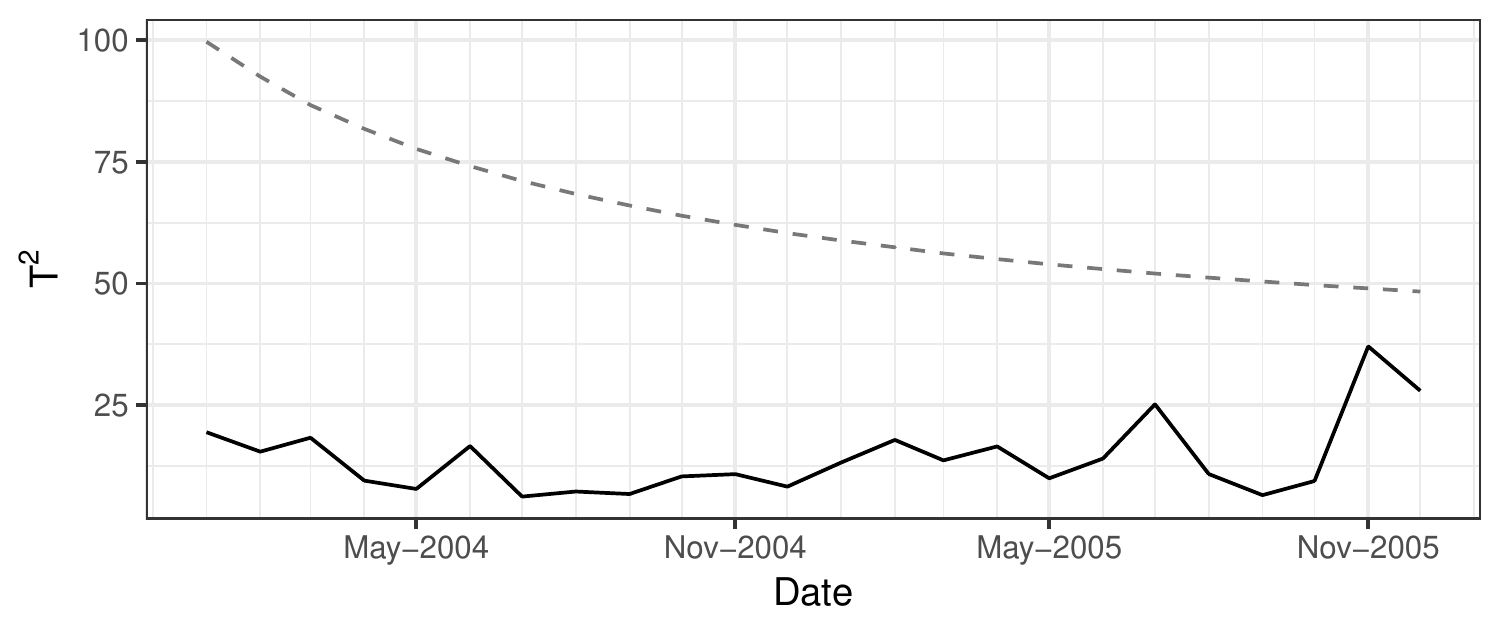}
  \caption{Hotelling's $T^2$ statistic (solid) and critical values
    (dashed), calculated monthly in the years 2004--2005 for the
    German meningococcal disease data.} \label{fig:hot2}
\end{figure}

We now calculate the Hotelling $T^2$ statistic for the (monthly) Meningococcal
data introduced in Section \ref{sec:motivation}.
Because this method assumes that the data follows a multivariate normal
distribution, it is at an immediate disadvantage when applied to case data with
low counts. We therefore aggregate this data, first over time to form
monthly counts for all 413 districts of Germany, and then over space to obtain
the corresponding time series for each of the country's 16 states, i.e.\ $p=16$.
For the purposes of illustration, rather than epidemiological correctness, we
also combine the cases across the two different finetypes B (MenB) and C (MenC).
In Figure \ref{fig:hot2}, we show the calculated $T^2$ statistics and the
critical values at significance level corresponding to an $\text{ARL}_0$ of
3 years. The years 2002--2003 were used as a baseline period for the estimation
of the mean vector and covariance matrix using the standard sample formulas
\citep[see e.g.][]{Rencher2012},
and these parameter estimates were updated based on \emph{all} available data at
each time step. Note that it may be advisable to use robust estimators of the
mean vector and covariance matrix in practice; we chose the standard estimators
here because parameter estimation is not the focus of this illustration.
\citet{Reinhardt2008} analyzed the meningococcal disease data
for the years 2004--2005 using the EpiScanGIS \citep{episcangis} software and
found one cluster in the period. For comparison, we therefore run our analysis
in the same time interval. Figure \ref{fig:hot2} shows the monthly time series
of the $T^2$ statistic (solid line), along with a critical value (dashed line).

\noindent
As can be seen in Figure \ref{fig:hot2}, Hotelling's method fails to detect
any outbreak at the chosen significance level (here $\alpha = 1/36$,
corresponding to one expected false detection every 3 years). An apparent flaw
is that the
detection threshold, which is based on an $F$ distribution with one of the
degrees of freedom parameters increasing with $n$, will decrease monotonically
with time, eventually reaching its asymptote. This happens regardless of
significance level, and it may therefore be useful to explore other ways of
obtaining detection thresholds---simulation-based, perhaps.
Another drawback, which is not exclusive to Hotelling's method, is that even if
a detection is made, the method does not inform us of \emph{where} (or which
variables) caused the threshold to be exceeded. Methods that remedy this problem
are the topic of the next section.

\subsection{Spatio-temporal Cluster Detection}
For some diseases, it may suffice to monitor the univariate time series of cases
aggregated across a country, using the methods described in Section
\ref{sec:univariate}. A detection signal may then prompt further investigation
into the time, location and cause of the potential outbreak.
For many diseases however, a large increase in the case count of a small
region can be drowned out by the noise in the counts of other
regions as they are combined. In order to detect emerging outbreaks of such
diseases before they grow large, the level of aggregation needs to be smaller,
perhaps even at the level of individual cases---a tradeoff between variability
and detecting a pattern.
Despite the possibilities of long-range transmission enabled by modern means of
transportation, transmission at a local level is still the dominant way
in which most infectious diseases spread \citep{Hohle2016}. Thus, the statistical
methods used to detect the
\emph{clusters} that arise should take into account the spatial
proximity of cases in addition to their proximity in time.
We begin this section by describing methods for \emph{area-referenced}
discrete-time data, in which
counts are aggregated by region and time period. This is followed by Section
\ref{sec:pointproc}, discussing methods for continuous space, continuous time data.

%
%
%
%
%


\subsubsection{Methods for Area-Referenced Data} \label{sec:area_referenced}
Aggregation by region and time period may in some cases be required by public
health authorities due to the way the data collection or reporting works, or for
privacy reasons. Legislation and
established data formats may thus determine at what granularity surveillance data
is available. For example, the
CASE surveillance system \citep{cakici_etal2010} used by the Public Health
Agency of Sweden is limited to data at county (\emph{län}) level, with the least
populous of the 21 counties
having a population of about 58,000 people.
The monitored data in the area-referenced setting is typically of the form $\{y_{it}\}$,
where $i=1,\ldots,N$ denotes the number of (spatial) regions, such as counties,
and the index $t=1,\ldots,T$ denotes time intervals of equal length.
Cluster detection in this setting involves identifying a set $Z$ of regions in which
a disease outbreak is believed to be emerging during the $D=1,2,\ldots$ most
recent time periods (weeks, for example). We will denote such a space-time \emph{window}
by $W$, and its complement by $\overline{W}$.
A well-established methodology for cluster detection for this task is
that of \emph{scan statistics}, which dates back to the 1960's with work done
by \citet{Naus1965}, and that took on its modern form after the seminal
papers by \citet{Kulldorff1995} and \citet{Kulldorff1997a}. These methods,
which were extended from the spatial to the spatio-temporal context by \citet{Kulldorff2001},
are widely used amongst public health departments thanks to
the free software \textsc{SaTScan}™ \citep{SaTScan}.

\paragraph{Example: Kulldorff's prospective scan statistic} \label{sec:kulldorff}
To illustrate the typical procedure of using a scan statistic designed for space-time
cluster detection, we work through
the steps of calculating and applying of the most popular such methods, often
simply referred to as \emph{Kulldorff's (prospective) scan statistic}
\citep{Kulldorff2001}. This method assumes that
the count $Y_{it}$ in region $i$ and time $t$ follows a Poisson distribution
with mean $q_{it} \cdot b_{it}$. Here, $b_{it}$ is an `expected count' or `baseline',
proportional to the population at risk in region $i$ at time $t$. For example,
these could be constrained such that the sum of the expected counts over all
regions and time periods equals the total of the observed counts during the
time period under study. That is, $\sum_{it} b_{it} = \sum_{it} y_{it}$.
The factor $q_{it} > 0$, often called the \emph{relative risk}, is assumed to be the
same $q_{it} = q$ for all $i$ and $t$ provided there is no outbreak. This
constitutes the null hypothesis. In the case
of an outbreak however, it is assumed that the relative risk is higher
inside a space-time window $W = Z \times \{T-D+1, \ldots, T\}$, consisting of a
subset of regions $Z \subset \{1,\ldots,N\}$ and stretching over the $D$ most
recent time periods. That is, for $i \in Z$ and $t > T - D$ we have
$\text{E}[Y_{it}] = q_{W} b_{it}$, while for $i \not\in Z$ or $t \leq T - D$
it holds that $\text{E}[Y_{it}] = q_{\overline{W}} b_{it}$ with $q_{W} > q_{\overline{W}}$.
Here, $\overline{W}$ is the complement of $W$. This multiplicative increase in
the baseline parameters inside a space-time window is typically how outbreaks
are modelled for scan statistics.
For scan statistics applied to prospective
surveillance, it is important to note that all potential space-time clusters
have a temporal duration that stretches from the most recent time period backwards,
without interuptions. This means that no inactive outbreaks are considered, and that
any space-time window included in an analysis can be thought of as a cylinder with
a base formed by the perimiter of the geographical regions covered by the window,
and a height equal to its length in time.

\paragraph{Definition and calculation of the scan statistic} \label{sec:scan_definition}
Of course, there could be many space-time windows $W$
for which the alternative hypothesis is true. If one would conduct hypothesis
tests for each window separately---and the number of such windows could well
be in the thousands or hundreds of thousands in typical applications---this would
result in a very large number of false positives for standard significance
levels such as 0.05. This problem could be counteracted by lowering the nominal
significance level to a miniscule value, or by using some other repeated testing
strategy, but this would in turn allow only very large outbreaks (in terms of
$q_{W}$ relative to $q_{\overline{W}}$) to be captured. The solution
proposed by \citet{Kulldorff2001} is to focus only on the window $W$ that stands out
compared to the others, as measured by the size of likelihood ratio statistics
calculated for all windows $W$ of interest. By calculating the maximum of all
such statistics, and using the distribution of this maximum to calculate $P$-values,
the `most anomalous' space-time cluster can be identified.
To calculate a likelihood ratio for a given space-time window
$W = Z \times \{1,2,\ldots,D\}$, one must first calculate the maximum
likelihood estimates of the relative risks $q_W$ and $q_{\overline{W}}$.
For Kulldorff's Poisson scan statistic, these are easily computed as
\begin{align} \label{eq:Kulldorff_q}
\hat{q}_{W} = \frac{Y_W}{B_W},
\quad \hat{q}_{\overline{W}} = \frac{Y - Y_W}{Y - B_W}
                             = \frac{Y_{\overline{W}}}{B_{\overline{W}}},
\end{align}
where
\begin{align} \label{eq:Kulldorff_sums}
Y_W = \sum_{(i,t) \not\in W} y_{it},
B_W = \sum_{(i,t) \in W} b_{it}, \text{ and }
Y =  \sum_{i=1}^N \sum_{t=1}^T y_{it} = \sum_{i=1}^N \sum_{t=1}^T b_{it}.
\end{align}
Thus, the likelihood ratio statistic conditional on the window $W$ is then
given by
\begin{align} \label{eq:pbpoi}
  \lambda_W = \left( \frac{Y_W}{B_W} \right)^{\!\!Y_W} \!\!\!
            \left( \frac{Y - Y_W}{Y - B_W} \right)^{\!\!Y - Y_W} \!\!
            \IX_{\{Y_W > B_W\}},
\end{align}
up to a multiplicative constant not dependent on $q_W$ or $q_{\overline{W}}$.
Here, $\IX_{\{ \cdot \}}$ is the indicator function.
This statistic is then calculated for all space-time windows $W$ of interest,
and the \emph{scan statistic} is defined as the maximum of all such statistics:
$\lambda^* = \max_W \lambda_W$. The corresponding window $W^*$, often called the
\emph{most likely cluster} (MLC), is thus identified.

\paragraph{Hypothesis testing} \label{sec:scan_hypothesis}
Because the distribution of $\lambda^*$ cannot be derived
analytically, a Monte Carlo approach to hypothesis testing is often taken,
whereby new data for each region $i$ and time $t$ is simulated under the null
hypothesis using the expected counts $b_{it}$. For Kulldorff's scan statistic,
the sampling is made conditional on the total observed count $C$, leading to a
multinomial distribution over regions and time intervals for the new counts.
This sampling is repeated $R$ times, and for each sample $r$, a new scan statistic
$\lambda^*_r$ is computed. A Monte Carlo $P$-value for the observed scan statistic can then
be calculated in standard fashion using its rank amongst the simulated values:
\begin{align*}
  P = \frac{1 + \sum_{r=1}^R \IX\{ \lambda^*_r > \lambda^*_{\text{obs}} \}}{1 + R}.
\end{align*}
Typically, a number such as $R=999$ or $R=9999$ is used in order to get a fixed
number of digits for the $P$-value. For prospective surveillance, past analyses---and potentially
future ones---should also be accounted for when conducting hypothesis tests,
in order to avoid a greater number of expected false positives than implied by
the nominal significance value (i.e.\ a multiple testing problem). The solution suggested
by \citet{Kulldorff2001} is to expand the set of replicates $\{\lambda^*_r\}_{r=1}^R$ above
by including replicates calculated in past analyses. If too many past analyses
are included however, the hypothesis tests could become too conservative.
\citet{Kulldorff2001} therefore recommends including only the most recent $\ell$ analyses, where
$\ell$ could be chosen to achieve a certain false positive rate during a given
monitoring period, for example. This practice has been the subject of a heated debate
recently, with \citet{Correa2015a} asserting that any nominal significance level
$\alpha$ used for prospective surveillance with Kulldorff's scan statistic
is unrelated to the average run length (ARL) and recurrence interval (RI) which are
commonly used in prospective surveillance. Similar points have been raised
earlier by \citet{Woodall2006}, \citet{Joner2008}, and \citet{Han2010}.
In a rebuttal, \citet{Kulldorff2015} explain that in one of the
three prospective cases considered by \citet{Correa2015a}, the simulations performed
actually show the expected result, and that in the other two cases the concerns
raised are actually misunderstandings. \citet{Correa2015b} later clarify that
failure to account for future analyses remain a concern, despite the comments
by \citet{Kulldorff2015}. In the latest reply to the debate, \citet{Tango2016}
states that both of the previous parties are in fact wrong: \citet{Correa2015a}
for being unrealistic in their consideration of an indefinite number of future
analyses and for focusing on the ARL in a setting for which the spatial component
of outbreaks is at least as important as the temporal one (because the ARL does
not inform us of the spatial spreading of the disease), and
\citet{Kulldorff2001} and \citet{Kulldorff2015} for
performing a prospective analysis \emph{conditional} on the total number of
observed cases in any given study period. Rather, \citet{Tango2016} reiterates
the point made by \citet{Tango2011} that such an analysis should be
\emph{unconditional} on the total count. This last point will be expanded upon
below, but the overall implication for those wishing to apply scan statistics in
prospective settings is to carefully weigh the benefits and costs of setting the
threshold for ``sounding the alarm'' at a particular level.

Given that the number of space-time windows to be included in the
analysis can range in the thousands or hundreds of thousands, the calculation of
Monte Carlo $P$-values means an $R$-fold increase of an already
high computational cost. One way to reduce this computational cost is to calculate a smaller
number of Monte Carlo replicates of the scan statistic, fit an appropriate
distribution to these replicates, and then compute the tail probability of the
observed scan statistic from the fitted distribution. \citet{Abrams2010} tested
such a procedure for a number of different distributions (Gumbel, gamma, log-normal,
normal) on Kulldorff's scan statistic and others. The authors found that the
Gumbel distribution yielded approximate $P$-values that were highly accurate
in the far tails of the scan statistic distribution, in some scenarios making it
possible to achieve the same rejection power with one tenth as many Monte Carlo
replicates. Another possibility is to circumvent simulation
altogether by comparing the value of the scan statistic computed on the current
data to values calculated in the past, provided no outbreaks are believed have been
ongoing in the data used for these past calculations. \citet{Neill2009a} compared
this approach to Monte Carlo simulation with standard and Gumbel $P$-values, and
found that the latter two methods for calculating $P$-values gave misleading
results on three medical data sets, requiring a much lower significance level
than originally posited to reach an acceptable level of false positives.

\paragraph{Cluster construction} \label{sec:cluster_construction}
A second way of limiting the computational cost of running a scan statistic
analysis is to limit the search to clusters with a compact spatial component
(\emph{zone}) $Z$, in the sense that all regions
inside the cluster are close to one another geographically. This makes sense
for both computational and practical reasons, since many of the $2^N-1$ subsets of all
$N$ regions are spatially disconnected and therefore not of interest for detection of
diseases that emerge locally.
For instance, one could limit the search
to the $k$ nearest neighbors to each region $i=1,\ldots,N$, for $k=0,1, \ldots,K_{\text{max}}$,
where $K_{\text{max}}$ is some user-defined upper bound \citep[as in][]{Tango2011}, or
automatically chosen such that the largest zone with region $i$ as center
encompasses regions with a combined population that equals approximately 50\%
of the total population in the whole study area \citep[as in][]{Kulldorff2001}.
The set of zones obtained by these methods will be quite compact, which can be
problematic if the true outbreak zone consists of regions that all lie along a riverbank,
for example. To capture a richer set of zones, \citet{Tango2005} propose a
way to construct `flexibly shaped' zones, by considering all connected subsets of
the $K_{\text{max}}$ nearest neighbors of each region $i$, that still contain
region $i$ itself. This method can yield a vastly larger set of zones, but
becomes impractical in terms of run-time for $K_{\text{max}} > 30$ when the
number of regions is about 200 or more.

More data-driven approaches to finding the set of interesting zones can also be
taken. For example, \citet{Duczmal2004} consider the set of all connected
subgraphs as the set of allowable zones, and search the most promising clusters
using a simulated annealing approach. Here, the nodes in the graph searched are
the spatial regions, and edges exists between regions that share a common border.
Another more holistic approach is taken by \citet{Neill2013}, who propose a
framework in which regions are first sorted by priority based on observed counts
and estimated parameters, and the set of zones scanned then taken to be only the
top $n$ regions, for $n=1, 2, \ldots, N$. This method is discussed further in
Section \ref{sec:multivariate_scan}.

\paragraph{Illustration of Kulldorff's scan statistic} \label{sec:scan_illustration}
Once the set of clusters to be scanned have been determined, the analysis can
take place. Here, we apply Kulldorff's prospective scan statistic
\citep{Kulldorff2001} to the Meningococcal data considered earlier, now
aggregated to monthly counts for each of Germany's 413 districts (\emph{kreise}).
This scan statistic is implemented in the \textsc{R} package
\texttt{scanstatistics} \citep{scanstatistics} as the function
\texttt{scan\_pb\_poisson}.

\begin{figure}[H]
  \centering
  \includegraphics[scale=0.75]{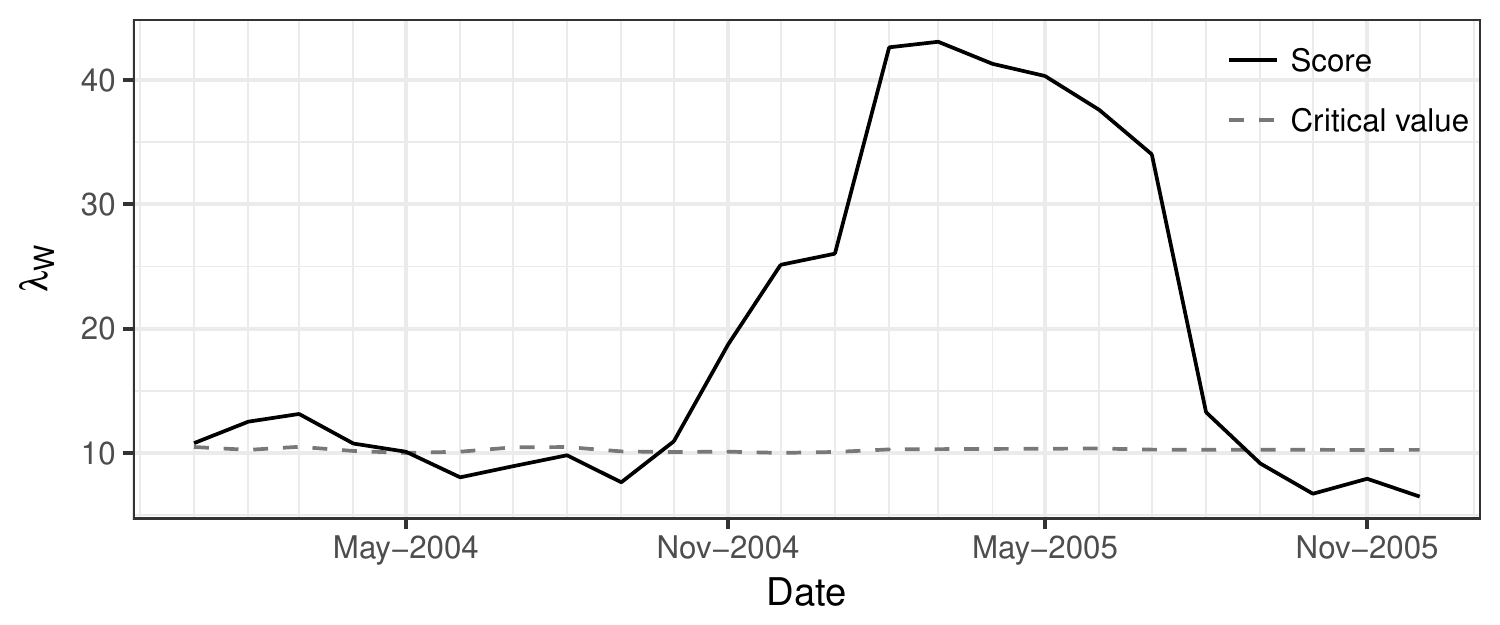}
  \caption{Observed value of Kulldorff's prospective scan statistic, calculated
    monthly in the years 2004--2005 for the German meningococcal disease
    data.} \label{fig:scan_score}
\end{figure}

\begin{figure}[H]
  \centering
  \includegraphics[scale=0.8]{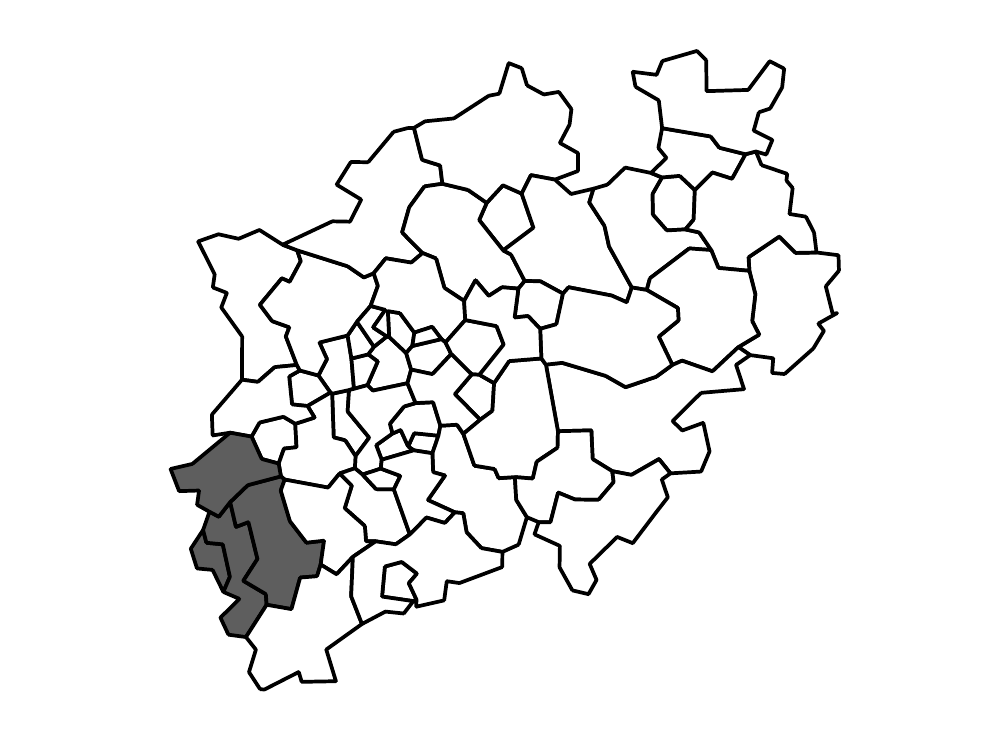}
  \caption{Districts of North Rhine-Westphalia, with the most likely cluster
           shaded in gray.} \label{fig:scan_mlc}
\end{figure}

In Figure \ref{fig:scan_score}, we show the resulting scan statistics for each
month of the study period (2004--2005). At each time step, the statistic was
calculated using at most the latest 6 months of data, and the $b_{it}$ for each
district and time point was estimated as
\begin{align}
  \hat{b}_{it} = \frac{Y}{T} \cdot \frac{\text{Pop}_i}{\text{Pop}_{\text{total}}}.
\end{align}
Here, $Y$ is the total observed count over all districts and time points, $T=6$
is the length of the study period, and $\text{Pop}_i$ and $\text{Pop}_{\text{total}}$
are the 2008 populations for district $i$ and all of Germany, respectively.
Critical values (sample quantiles) for the significance level $\alpha = 1/60$ were
obtained from Monte Carlo replication with $R = 99$ replicates, and previously
generated replicates were included in the calculation at each new time step.
\noindent
The most likely cluster detected by Kulldorff's prospective scan
statistic, for most months scanned, corresponds well to the region of highest
disease incidence in the years 2004--2005. The core cluster seems to be four
districts in North Rhine-Westphalia, one of them the city (urban district)
Aachen, and coincides with a confirmed cluster of Meningococcal disease
discussed in \citet{Meyer2012}. This cluster also matches that found by
\citet{Reinhardt2008}.
In Figure \ref{fig:scan_mlc}, we show a map of the counties of North
Rhine-Westphalia, with the detected cluster shaded in gray.

Figure \ref{fig:scan_score} shows that the given significance level
results in detection signals during much of the study period. This may not be
entirely plausible, and one could see this as an inadequacy of the Monte Carlo
method, as discussed in Section~\ref{sec:scan_hypothesis} (see in particular
the issues of debate mentioned there). More likely however, is that the Poisson
assumption of Kulldorff's prospective scan statistic is ill-suited for data with
such an abundance of zeros as the meningococcal data, considering that it
assumes a Poisson distribution for the counts. For this type of data, a scan
statistic based on e.g.\ the zero-inflated Poisson distribution
\citep[see][]{Cancado2014,Allevius2017} may perform better.

\paragraph{Developments in space-time scan statistics} \label{sec:developments}
Kulldorff's \citeyear{Kulldorff2001} prospective scan statistic was introduced
above to present the general procedure of using a scan statistic to detect
space-time disease clusters. Of course, many more such methods have
been devised since 2001, many to deal with the purported ``flaws'' of Kulldorff's
statistic. One such drawback of Kulldorff's prospective scan
statistic is that it requires data on the population at risk and its spatial
(and possibly temporal) distribution over the study area. The notion of a population
at risk may not be applicable in cases where the surveillance data consists of
counts of emergency department visists and over-the-counter medicine sales, for
example.
Noting this fact, \citet{Kulldorff2005} formulate a \emph{space-time permutation scan statistic},
which uses only the observed counts in the current study period and area to
estimate the baselines. Under the assumption that the probability of a case
occurring in region $i$, given that it occurred at time $t$, is the same for all
times $t=1,\ldots,T$, the baseline (expected value) for the count in region $i$ at time $t$
is estimated by \citet{Kulldorff2005} as
\begin{align} \label{eq:b_estimate}
b_{it} = \frac{1}{Y} \left( \sum_{j=1}^N y_{jt} \right) \left( \sum_{\tau=1}^T y_{i \tau} \right).
\end{align}
The analysis is thus conditional on
the marginal counts over regions and time points, leading to a hypergeometric
distribution for the total count inside a space-time cluster $W$. This distribution
can in turn be approximated by a Poisson distribution when the marginal counts
inside the cluster are small compared to the overall total count $C$. Thus,
a likelihood ratio statistic can be calculated using Equation \eqref{eq:pbpoi},
leading to a scan statistic of the same form as \citet{Kulldorff2001}. For hypothesis testing
however, the random sampling is done such that the marginal counts over regions
and time points are preserved, rather than just the total count.

The prospective \citep{Kulldorff2001} and space-time permutation
\citep{Kulldorff2005} scan statistics both conduct hypothesis testing using expected
counts that are \emph{conditional} on the total or marginal counts in the observed
data. This has been met with some critique, particularly from \citet{Tango2011}
who give a simple (albeit extreme) example showing that if counts are increased
uniformly in all regions under surveillance, these types of scan statistics will
fail to detect that something out of the ordinary has happened. Using less
extreme circumstances, \citet{Neill2009a} demonstrate that a `conditional' scan
statistic of this sort has low power to detect outbreaks that affect a large
share of the regions under surveillance. When an outbreak actually is detected
in such a scenario, that scan statistic takes longer to do so than the
\emph{expectation-based} scan statistics used for comparison.
Introduced for Poisson-distributed counts by \citet{Neill2005} and
\citet{Neill2006a}, these scan statistics do not
condition on the observed total count in their analysis. Neither do they use the
most recent data for baseline parameter estimation; rather they calculate the baselines
$b_{it}$ and other non-outbreak parameters based on what we can \emph{expect} to
see from past data. In that sense, the analysis is split into two independent parts:
First, parameters of the distribution are
estimated on historical data believed to contain no outbreaks. This can be done
by any method regression preferable; typically by fitting a GLM \citep{GLMbook}
or a moving average to the data.
Second, the estimated parameters are used for simulating new counts from the
given probability distribution, and plugged into the calculation of the scan
statistic on both the observed and simulated data in order to conduct Monte Carlo
hypothesis testing.
Expectation-based scan statistics have been formulated for the Poisson
\citep{Neill2005,Neill2009b}, Gaussian \citep{Neill2006a}, negative binomial
\citep{Tango2011}, and zero-inflated Poisson \citep{Allevius2017} distributions,
among others.

The drawback of these scan statistics, as well as those mentioned at the
beginning of this section, is
that they scan over a fixed set of space-time windows $W$. If the true outbreak
cluster is not among the windows scanned, the outbreak cannot be exactly identified.
As also mentioned, the number of windows to be scanned poses a computational burden,
particularly if Monte Carlo hypothesis is to be conducted. To overcome these issues,
\citet{Neill2012} proposes a `Fast Subset Scan' framework for making fast searches
for the top-scoring cluster, unconstrained by any pre-defined set of windows to
be scanned. In particular, \citeauthor{Neill2012}
introduces a `Linear Time Subset Scanning' (LTSS) property, which is shown to hold for several
members of the exponential family of distributions (or rather, scan statistics based
thereon). For scan statistics with this property, a \emph{score function} (such as the
likelihood ratio in Equation \eqref{eq:pbpoi}) is paired with a \emph{priority function},
defined e.g.\ as the ratio of of count to baseline.
The latter function is used to sort the data in order of priority, after which
the former function only needs to be applied to increasing subsets of the
ordered records. This allows an unconstrained search for the top-scoring subset
in linear time (plus the cost of sorting the data records); the method can also
be modified to search only for clusters fulfilling spatial constraints, such
as subsets of the $K$ nearest neighbors of each region. The Fast Subset Scan
framework can also be extended to multivariate space-time data, as covered next.


\paragraph{Multivariate scan statistics} \label{sec:multivariate_scan}
Until now, the scan statistics described have been univariate in the sense that
the counts $\{y_{it}\}$ typically consist of cases of a single disease or symptoms
thereof, each count having both spatial and temporal attributes.
In practice however, public
health authorities monitor a multitude of such data streams simultaneously. If
each analysis is done without regard for the others, this will undoubtedly yield
a number of false positives higher than desirable. Alternatively, if the significance
level of each analysis is adjusted for the others, the power to detect an outbreak
in any of the data streams diminishes. These issues can become by analyzing all
data streams \emph{jointly}, an endeavour that can be particularly fruitful when
the streams pertain to the same phenomenon to be detected; typically a disease
outbreak. For example, some diseases have multiple symptoms and patients may
seek help in different ways. A simultaneous increase in over-the-counter flu
medication sales at pharmacies and respiratory symptoms reported at a local
clinic may in such a case be indicative of an influenza outbreak, but this signal
of an outbreak may be missed if each of the two data streams are considered individually.

With this motivation and inspired by an idea of \citet{Burkom2003},
\citet{Kulldorff2007} formulate a multivariate scan
statistic based on Kulldorff's prospective scan statistic \citep{Kulldorff2001}.
The data in this setting can be represented as a collection of vector counts
$\{\bs{y}_{it}\}$, where $\bs{y} = (y_{it}^{(1)}, \ldots, y_{it}^{(M)})'$ are the
counts for each of the $M$ data streams monitored.
The scan statistic is calculated by first processing each data stream separately,
calculating a likelihood ratio statistic using Equation \eqref{eq:pbpoi}
for each space-time window $W$, just as is done with the univariate scan statistic.
For those windows whose aggregated count exceeds the aggregated baseline,
the logarithm of these likelihood ratios are added to form a statistic for the
window as a whole. The scan statistic is then defined as the maximum of all
such statistics, so that the most likely cluster can be identified as the
regions, time intervals \emph{and} data streams making a positive contribution
to the maximum statistic.

Building upon the Fast Subset Scan framework \citep{Neill2012} cited earlier,
\citet{Neill2013} present two computationally efficient ways to detect clusters
in space-time data, with and without spatial constraints on these clusters. The
first method, \emph{Subset Aggregation}, is an extension of the work done by
\citet{Burkom2003}. It assumes that if an outbreak occurs, it has a multiplicative
effect on the baselines $b_{it}^{m}$ that is equal across all regions, time
points and data streams affected by the outbreak. That is,
for those regions, times and streams $(i,t,m)$ affected by the outbreak, the
expected value of the count $y_{it}^{(m)}$ is $q b_{it}^{(m)}$ rather than
$b_{it}^{(m)}$.
This allows the counts and baselines within each \emph{subset} (cluster) to be
\emph{aggregated}, so that the Subset Aggregation scan statistic can be reduced
to a univariate scan statistic for the cluster. The second method is the one
previously proposed by \citet{Kulldorff2007}, in which an outbreak affects
each data stream separately through a stream-specific multiplicative factor
$q_m$. \citet{Neill2013} then demonstrates how these methods can be combined
with the Fast Subset Scan framework for scan statistics satsifying the LTSS
property \citep{Neill2012}, yielding fast, exact detection algorithms when
either the number of data streams or the number of regions are small, and
fast approximate (randomized) algorithms when there are too many regions and
streams for all subsets of each to be scanned. If hypothesis testing is to take
place, this can be done using Monte Carlo replication as described earlier.
Again, such replication can come at a high computational cost, and some efforts
have therefore been made to avoid it altogether.

\paragraph{Bayesian scan statistics}
\citet{Neill2006} introduce the \emph{Bayesian Spatial scan statistic} for cluster
detection, based on Kulldorff's \citeyear{Kulldorff1997a} original
scan statistic. The method is easily extended to a spatio-temporal setting,
which is that described below. In
\citeauthor{Kulldorff2001}'s \citeyear{Kulldorff2001} model,
the data $\{y_{it}\}$ are assumed to be Poisson distributed with expected values
$q \cdot b_{it}$, the relative risk $q$
varying depending on whether an outbreak is ongoing or not, and estimated by
maximum likelihood. In the model of \citet{Neill2006}, the parameters $q$ are
instead given conjugate gamma distribution priors, with prior probabilities
tuned to match the occurrence of an outbreak in each possible outbreak cluster
considered. With the conjugate
prior for the relative risks, and baselines $\{b_{it}\}$ estimated from historical
data, simple analytical formulas can be derived for the marginal probabilities
of the data (relative risks integrated out), in the end resulting in the posterior
probability of an outbreak for each cluster considered, and for the non-occurrence of
an outbreak. Thus, no Monte Carlo replications need to be made.

To examplify the Bayesian scan statistic \citep{Neill2006}, suppose the
null hypothesis of no outbreak states that each count is distributed as a
Poisson random variable with mean $q \cdot b_{it}$, where $b_{it}$ is fixed and
$q \sim \text{Gamma}(\alpha_{\text{all}}, \beta_{\text{all}})$. After
marginalizing over the distribution of $q$, the likelihood under the null
hypothesis becomes the negative binomial (a.k.a.\ gamma-Poisson mixture)
probability mass function:
\begin{align} \label{eq:scanbayes_null}
 \P(\bs{y} | H_0) &= \frac{\Gamma(\alpha_{\text{all}} + Y)}
                    {Y! \, \Gamma(\alpha_{\text{all}})}
               \left(
                 \frac{\beta_{\text{all}}}{\beta_{\text{all}} + B}
               \right)^{\alpha_{\text{all}}}
               \left(
                 \frac{B}{\beta_{\text{all}} + B}
               \right)^{\alpha_{\text{all}}},
\end{align}
where $\bs{y}$ represents the entire data set, $Y$ is the sum of all counts,
and $B$ the sum of all baselines $b_{it}$.
The alternative hypothesis states that an outbreak is occuring in a space-time
window $W$, with a prior distribution placed over all potential windows $W$.
For a given $W$, it is assumed that counts inside $W$ have a relative risk
$q$ distributed as $q \sim \text{Gamma}(\alpha_W, \beta_W)$,
while those counts outside have a corresponding distribution for $q$
with parameters $\alpha_{\overline{W}}$ and $\beta_{\overline{W}}$. After
marginalizing over the relative risk distributions, the likelihood becomes
\begin{align}
\begin{split}
 \P(\bs{y} | H_1(W)) &= \frac{\Gamma(\alpha_W + Y_W)}
                             {Y_W! \, \Gamma(\alpha_W)}
                   \left(
                     \frac{\beta_W}{\beta_W + B_W}
                   \right)^{\alpha_W}
                   \left(
                     \frac{B_W}{\beta_W + B_W}
                   \right)^{\alpha_W} \\
                   &\times
                   \frac{\Gamma(\alpha_{\overline{W}} + Y_{\overline{W}})}
                    {Y_{\overline{W}}! \, \Gamma(\alpha_{\overline{W}})}
                   \left(
                     \frac{\beta_{\overline{W}}}{\beta_{\overline{W}} + B_{\overline{W}}}
                   \right)^{\alpha_{\overline{W}}}
                   \left(
                     \frac{B_{\overline{W}}}{\beta_{\overline{W}} + B_{\overline{W}}}
                   \right)^{\alpha_{\overline{W}}},
\end{split}
\end{align}
where $Y_W$ and $B_W$ is the sum of counts and baselines inside $W$,
respectively, $Y_{\overline{W}} = Y - Y_W$, and $B_{\overline{W}} = B - B_W$.
With a prior $\P(H_0)$ placed on the null hypothesis, and similarly
$\P(H_1(W))$ for each outbreak scenario, one obtains the posterior probabilities
\begin{align}
  \P(H_1(W) | \bs{y}) &= \frac{ \P(\bs{y} | H_1(W)) \P(H_1(W))}{\P(\bs{y})}, \\
  \P(H_0 | \bs{y}) &= \frac{ \P(\bs{y} | H_0) \P(H_0)}{\P(\bs{y})},
\end{align}
where $\P(\bs{y}) = \P(\bs{y} | H_0) \P(H_0) + \sum_W \P(\bs{y} | H_1(W)) \P(H_1(W))$.
\citet{Neill2006} gives advice for eliciting the priors $\P(H_0)$ and $\P(H_1(W))$,
and also for specification of the hyperparameters of each relative risk
distribution.

\begin{figure}[H]
  \centering
  \includegraphics[scale=0.75]{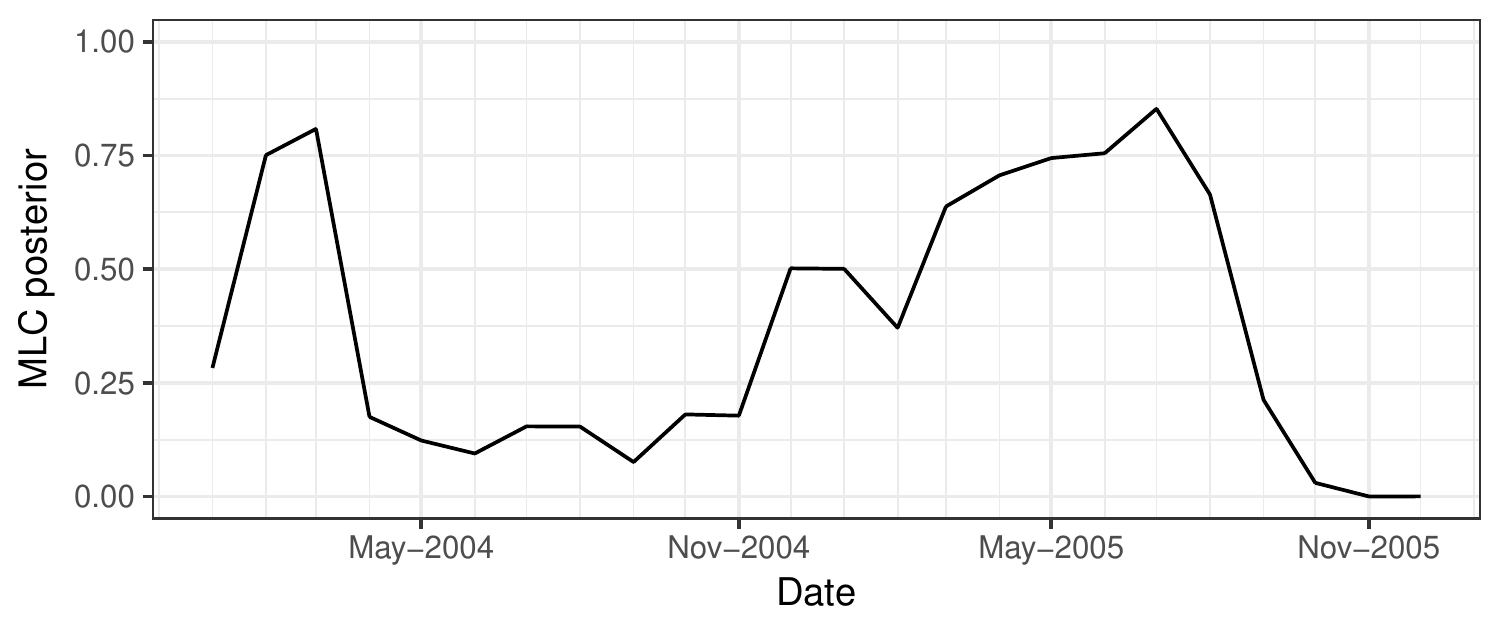}
  \caption{The posterior outbreak probability $\P(H_1(W)|\bs{y})$ for the most
    likely cluster (MLC) $W$ detected in each month of the analysis} \label{fig:bayMLCprob}
\end{figure}

The space-time extension of the Bayesian spatial scan statistic
\citep{Neill2006} is available as the function \texttt{scan\_bayes\_negbin} in
the \texttt{scanstatistics} R package \citep{scanstatistics}. To illustrate, we
run this scan statistic on the same data as in the illustration of Kulldorff's
scan statistic
above. As hyperparameters,
we set $b_{it}$ to the estimate in Equation \eqref{eq:b_estimate}, and let
all gamma distribution parameters $\alpha_{(\cdot)}$ and $\beta_{(\cdot)}$ be equal
to 1. The exception is $\alpha_W$, which we assume to be the same for all $W$.
We give this parameter a discrete uniform prior on equally spaced values between
1 and 15 in the first month, and let subsequent months use the posterior
distribution from the previous month as a prior. We also set
$\P(H_1) = 1 - \P(H_0) = 10^{-7}$, which is on the order of magnitude of the
incidence of meningococcal disease in Germany in 2002--2008.
In Figure \ref{fig:bayMLCprob}, we show the posterior outbreak probability
$\P(H_1(W)|\bs{y})$ at each month of the analysis, for the space-time window $W$
which maximizes this probability.
\noindent
Figure \ref{fig:bayMLCprob} shows clear similarities to the scores calculated
for Kulldorff's scan statistic, shown in Figure \ref{fig:scan_score}. Further,
half of all most likely clusters reported by each method are the same. The
difference, as discussed earlier, is that the output of the Bayesian scan
statistic is a posterior probability rather than the maximum of a likelihood
ratio

The Bayesian scan statistic was later extended to a multivariate setting by
\citet{Neill2010}. This \emph{Multivariate Bayesian Scan Statistic} (MBSS)
is also capable of detecting multiple event types, thus allowing it, for example,
to assign probabilities to outbreaks of different diseases based on records of
symptom data. A downside of this method is that the set of clusters to be searched
must be specified, and the prior probability of an outbreak in each is (typically)
uniform over all clusters, regardless of size and shape. To remedy this defect,
\citet{Neill2011} modify the MBSS by specifying a hierarchical spatial prior over all
subsets of regions to be scanned. This method is shown to be superior to the
MBSS in spatial accuracy and detection timeliness, yet remaining computationally
efficient. More recently, for univariate data, a Baysian scan statistic based on
the zero-inflated Poisson distribution has been proposed by \citet{Cancado2017}.

\paragraph{Alternative methods for cluster detection in area-referenced data}
Scan statistics are not the only option for cluster detection of the sort discussed
above. To give an example, there is the \emph{What's Strange About Recent
Events?} (WSARE) \citep{Wong2005} method, available as software, which can detect
outbreaks in multivariate space-time data sets by using a rule-based technique
to compare the observed data against a baseline distribution. WSARE uses
Bayesian networks, association rules and a number of other techniques to produce
a robust detection algorithm. Other examples with more detail than space allows
for here can be found e.g.\ in \citet{Rogerson2008}.

\subsubsection{Methods for Point Process Data} \label{sec:pointproc}
The methods discussed in the previous section were applicable to data which has
been aggregated over space and time. While such an accumulation may be the
inherent form of the data, the loss
of granularity could impede both the timeliness and spatial accuracy with which
outbreaks are detected. When data with exact coordinates and time stamps are
available, it may thus be beneficial to analyze data in this format. In this section we
therefore assume that the available data is of the form $\{\bs{s}_i, t_i\}_{i=1}^n$,
where $\bs{s} = (x,y)$ are the coordinates (longitude and latitude) of the event (disease case, typically),
and $t$ the time of occurrence. We assume an ordering $t_1 < t_2 < \ldots < t_n$,
and that the study region is defined by the study area $\mathcal{A}$ and the
surveillance timer interval $(0, \mathcal{T}]$.

One starting point to analyze such data is to adapt a purely temporal surveillance
method to a spatio-temporal setting. \citet{Assuncao2009} does so by combining
a non-homogeneous Poisson point process partially observed on $\mathcal{A} \times (0, \mathcal{T}]$
with the Shirayev-Roberts (SR) statistic \citep{Shirayev1963,Roberts1966,Kennet1996},
utilizing the martingale property of the latter to establish a protocol for
achieving a desired ARL with minimal parameter input by the user.
Aside from the ARL, the user needs to specify a radius $\rho$ defining the
maximum spatial extent of the outbreak cluster, as well as a parameter
$\epsilon > 0$ which measures the relative change in the density within the outbreak cluster
as compared to the non-outbreak situation. The SR statistic in
\citeauthor{Assuncao2009}'s \citeyearpar{Assuncao2009} formulation is then defined
as
\begin{align} \label{eq:SR}
  R_n &= \sum_{k=1}^n \Lambda_{k,n}, \text{where} \\
  \Lambda_{k,n} &= (1 + \epsilon)^{N(Y_{k,n})} I_{Y_{k,n}}(x_y, y_i, t_i)
                   \exp \left( -\epsilon \mu(Y_{k,n})  \right).
\end{align}
Here, $Y_{k,n}$ is the cylinder defined by the ball of radius $\rho$ centered
on the location of the $k$th event and the time interval $(t_k, t_n]$ between the
$k$th and the $n$th event, $N(Y_{k,n})$ is the number of events inside that cylinder,
$\mu(Y_{k,n})$ is the expected number of events inside the cylinder if there
is no space-time clustering, and $I_{Y_{k,n}}$ is an indicator taking the value
1 if the $k$th event is inside $Y_{k,n}$ and 0 otherwise.
The outbreak signal goes off if $R_n$ is larger than the specified ARL, and the
outbreak cluster is identified as the cylinder $Y_{k,n}$ for which $\Lambda_{k,n}$
is maximized.
The quantity $\mu(Y_{k,n})$ requires the specification of
two densities related to the process' intensity function,
but since this is too much to ask from the user, $\mu(Y_{k,n})$ is instead
replaced by the estimate
\begin{align}
  \hat{\mu}(Y_{k,n}) = \frac{N\left( B(\bs{s}_k, \rho) \times (0, t_n] \right)
                        \cdot     N\left( \mathcal{A} \times (t_k, t_n] \right)}
                   {n},
\end{align}
i.e.\ the product of the number of events within the disk $B(\bs{s}_k, \rho)$,
regardless of time of occurrence, and the number of events that occurred
in the time interval $(t_k, t_n]$ anywhere within the whole study region,
divided by the total number of events. Every new incoming event thus requires
the re-calculation of all terms in Equation \eqref{eq:SR},
which \citet{Assuncao2009} demonstrate can be done in an efficient iterative
procedure.

\citeauthor{Assuncao2009}'s \citeyearpar{Assuncao2009} method was later
extended by \citet{Veloso2013} to handle the detection of multiple space-time
clusters. This is accomplished by (randomly) deleting excess events inside the
detected clusters and re-running the method with a new ARL threshold corrected
for the event deletion.

There have also been attempts to adapt retrospective methods for detection of
space-time point event clusters to a prospective setting. For example,
\citet{Rogerson2001} formulate
a local version of the (retrospective) Knox statistic \citep{Knox1964} and
combine it with a CUSUM framework to make the method suitable for prospective surveillance.
However, \citet{Marshall2007} later concluded that it is certainly not, owing
to a number of factors that strongly influence the performance of the method,
and which a user has little power to regulate properly. A remedy is offered by
\citet{Piroutek2014}, who redefine the local Knox statistic by \citet{Rogerson2001}
to be prospective rather than retrospective. For a given observation indexed by $i$,
the local Knox statistic $n_{st}(i)$ is defined as the number of observations that are closer
than $t$ units of time to $t_i$, and whose coordinates are less than a distance
$s$ away from $(x_i, y_i)$. In \citet{Rogerson2001}, the closeness in the temporal
dimension is measured in both directions, so that a $n_{st}(i)$ counts nearby events both
before and \emph{after} the $i$th event. This account of future events is included
in the CUSUM chart, which is not appropriate. \citet{Piroutek2014} instead let only
\emph{past} events enter into the calculation of $n_{st}(i)$, which turns out to
vastly improve the performance of the method in a prospective setting. As noted
by \citet{Marshall2007}, there are also problems with the normal distribution
approximation made to the statistic calculated for the CUSUM chart. Namely,
it can be very poor if the thresholds $s$ and $t$ are small, and this in turn
makes it difficult to set the appropriate control limit to achieve a given ARL.
\citet{Piroutek2014} instead propose to use a randomization procedure using past
data to establish the correct control limit, obviating the need of an approximate
distribution.

\citet{Paiva2015} later combines the modified local Knox statistic by
\citet{Piroutek2014} with the Cumulative Surface method proposed by \citet{Simoes2005},
allowing for a visualization of clusters in three dimensions.
In their method, the local Knox score of each event is smeared using a bivariate Gaussian
kernel function, and a threshold for detection is defined through the distribution
of the stochastic surfaces formed. In all, the method requires few parameters
as input from users, and needs no information of the population at risk.

\paragraph{Illustration of Assunção and Correa's SR statistic}
We now illustrate the method by \citet{Assuncao2009} on the meningococcal data
(finetype B only) considered earlier, this time using the coordinates and time
stamps of each event, rather than an aggregated count. The method is implemented
as the function \texttt{stcd} in the \textsc{R} package \texttt{surveillance}
\citep{Salmon2016a}.
As a validation of the method, we run the analysis for approximately the same
period as \citet{Reinhardt2008}, hoping that the results will be similar.
We guide our choice of the parameter $\rho$ required by this method by the
analysis of the meningococcal data by \citet{Meyer2012}; Figure 3 of this paper
suggests that setting $\rho = 75$ km is adequate. For the choice of the
parameter $\epsilon$, which is the relative change of event-intensity within the
to-be-detected cluster, we are guided by the simulation study by
\citet{Assuncao2009} and set $\epsilon = 0.2$. Lastly, we set the desired ARL to
30 days.

\begin{figure}[H]
  \centering
  \includegraphics[scale=0.9]{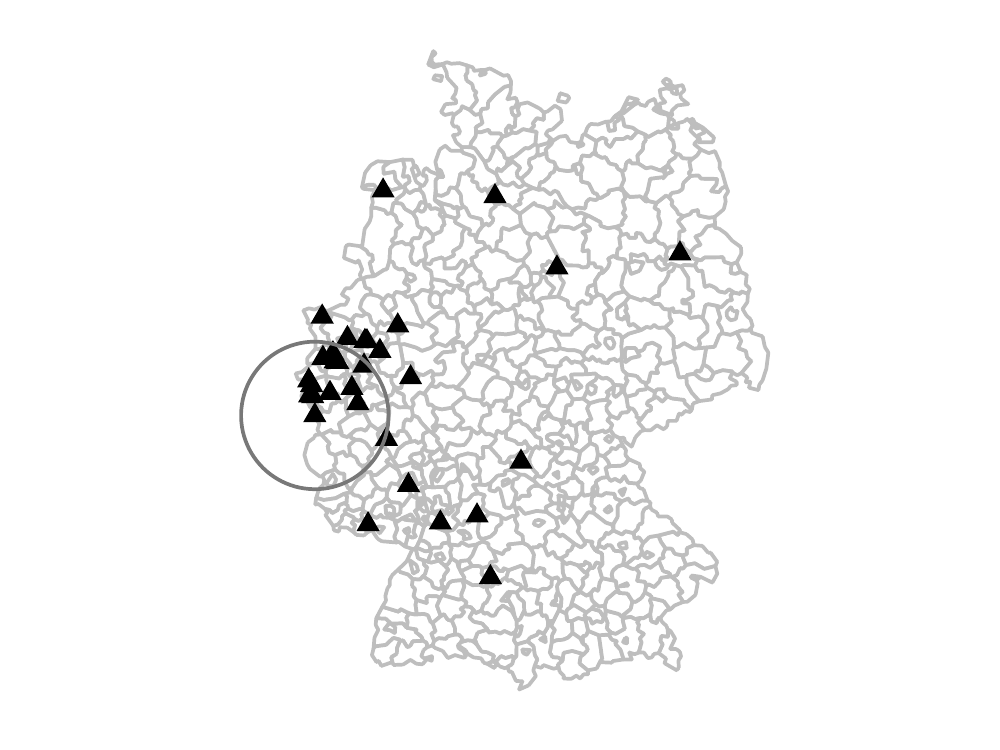}
  \caption{Detected cluster (gray circle) and observed disease events for the
    cluster detection method by \citet{Assuncao2009}.} \label{fig:stcd_map}
\end{figure}

In Figure \ref{fig:stcd_map}, the detected cluster is shown, with all events
up until the point of detection marked as triangles on the map.
\noindent
The cluster detected by the \texttt{stcd} function is centered on the city of
Aachen in the state North Rhine-Westphalia, which corresponds well to the cluster
marked in Figure 3 of \citet{Reinhardt2008}, and the dates also appear similar.
The cluster was detected only one day after its estimated start date, showing the
potential of \citeauthor{Assuncao2009}'s method in terms of timeliness.
All in all, the ability to use the spatial and temporal
attributes of each individual event for the detection of clusters is an
attractive feature when greater exactness in the origins and spatial extent of
outbreaks is important.

\section{Summary and Outlook}

This chapter presented temporal and spatio-temporal statistical
methods for the prospective detection of outbreaks in routine
surveillance data. Such data-driven algorithms operate at the
intersection of statistical methodology, data mining and computational
(big) data crunching. Simple methods have their virtues and speed of
implementation can be of importance; however, this should never be an
excuse for ignoring statistical facts when counts get small and
sparse. Facts which might surprise the non-statisticians uncomfortable
beyond the normal distribution.

Despite the many advances in the last 30 years, outbreak detection
algorithms will never replace traditional epidemiological
alertness. Nevertheless, they offer support for using the digital
epidemiologist's time more effectively. From a systems perspective,
however, it is our impression that detection algorithms have had
limited impact on practical outbreak detection so far. One reason is
that, historically, many algorithmic suggestions were not supported by
accessible software implementations. Another reason is that their
their usefulness is questioned by the many false alarms and a
misalignment between the users' needs and presentation of the alarms
found by the system. Statisticians---as part of interdisciplinary
teams---need to worry more about how to roll-out the proposed
algorithms in the public health organizations. As a step in this
direction, all detection methods presented in this chapter are readily
available from open-source packages in R. In particular, the
\texttt{surveillance} and \texttt{scanstatistics} packages\footnote{The R code
producing the results and graphics of this manuscript is available from
\url{https://github.com/BenjaK/ProspectiveDetectionChapter}}. Examples of
software supporting national surveillance systems by increased user
focus are the Swedish CASE system~\citep{cakici_etal2010} and the
German avoid@RKI~\citep{salmon_etal2016b}.\\

\subsubsection*{Acknowledgements}
Benjamin Allévius was supported by grant 2013:05204 by the Swedish
Research Council. Michael Höhle was supported by grant 2015-05182\_VR
by the Swedish Research Council. 

The authors thank Renato Assun{\c{c}}{\~{a}}o and Leonhard Held for their 
insightful feedback during the writing of the chapter.

%